\documentclass[runningheads]{svmult}

\usepackage{makeidx}   
\usepackage{graphicx}  
\usepackage{subeqnar}  
\usepackage{multicol}  
\usepackage{physprbb}  

\newcommand{\mnras}{MNRAS}
\newcommand{\aj}{AJ}
\newcommand{\apj}{ApJ}
\newcommand{\apjs}{ApJS}
\newcommand{\apjl}{ApJL}

\newcommand{\aap}{A\&A}

\newcommand{\pasp}{PASP}
\newcommand{\pasj}{PASJ}

\newcommand{\kms}{km\,s$^{-1}$}

\begin{document}
\title*{Doppler Tomography}
\toctitle{Doppler Tomography}
%
%
\titlerunning{Doppler Tomography}
%
\author{T.R.\ Marsh}
\authorrunning{T.R.\ Marsh}
%
%
\institute{Department of Physics and Astronomy, 
Southampton University, Highfield, Southampton SO17 1BJ}

\maketitle              

\begin{abstract}
I review the method of \index{Doppler tomography}Doppler tomography which
translates binary-star line profiles taken at a series of orbital
phases into a distribution of emission over the binary. I begin with a
discussion of the basic principles behind \index{Doppler tomography}Doppler tomography,
including a comparison of the relative merits of maximum \index{entropy}entropy
regularisation versus \index{filtering}filtered \index{back projection}back-projection for implementing the
inversion.  Following this I discuss the issue of noise in Doppler
images and possible methods for coping with it. Then I move on to look
at the results of Doppler Tomography applied to \index{cataclysmic variable}cataclysmic variable
stars. Outstanding successes to date are the discovery of two-arm
\index{spiral arms}spiral shocks in \index{cataclysmic variable}cataclysmic variable accretion discs and the probing
of the stream/magnetospheric interaction in magnetic cataclysmic
variable stars. \index{Doppler tomography}Doppler tomography has also told us much about the
stream/disc interaction in non-magnetic systems and the
\index{irradiation}irradiation of the \index{secondary star}secondary star in all systems. The latter indirectly
reveals such effects as \index{shadowing}shadowing by the accretion disc or
stream. I discuss all of these and finish with some musings on
possible future directions for the method. At the end I include a tabulation
of \index{Doppler map}Doppler maps published in refereed journals.
\end{abstract}

\section{Introduction}
Many rapidly rotating single and binary stars change little
during the course of a single rotation or orbit. The spots on
single stars can persist for many days, while \index{cataclysmic variable}cataclysmic variable
stars may stay in \index{outburst}outburst for over 100 orbits and in \index{quiescence}quiescence for
ten times longer still. However, for the observer, orbital rotation can 
cause considerable \index{variability}variability both in flux and spectra. This arises from 
a combination of changes in aspect angle and
\index{visibility}visibility, caused by geometrical effects, and the rotation of all
velocity vectors with the binary orbit. These effects are a blessing and 
a curse: without them we would know considerably less than we do about 
such stars, however the complex \index{variability}variability can be hard to interpret.

The method of \index{Doppler tomography}Doppler tomography was developed to unravel the emission
line variations of \index{cataclysmic variable}cataclysmic variable stars (CVs)
[48].  CVs are short period binary stars, with orbital
periods typically between $1.5$ and $10$ hours, which are beautifully
set up to allow us to study accretion. The stellar components of
the binary, a \index{white dwarf}white dwarf and a low-mass main-sequence star, are
faint, and their semi-detached configuration means that the geometry
is entirely specified by the \index{mass ratio}mass ratio and orbital \index{inclination}inclination
alone. Unfortunately, CVs are far too small to be resolved directly --
they typically subtend $< 10^{-4}$ seconds of arc at Earth -- and we
can learn nothing of their structure from direct imaging. Instead we
must turn to more indirect methods.  Two key methods are ``eclipse
mapping'', introduced by Horne [39] and reviewed in this
volume by Baptista, and ``\index{Doppler tomography}Doppler tomography'', the subject of this
chapter. Eclipse mapping relies on the geometrical information
contained in eclipse light curves; Doppler Tomography uses the
velocity information contained in Doppler-shifted light curves.

In this paper I detail the principles behind \index{Doppler tomography}Doppler tomography and
hope to give the reader a full picture of what is now a widely-applied
tool in the field. I follow the fundamentals with a short section on
accounting for stochastic noise, which has usually been ignored to
date, before finally moving onto a survey of results.  Although I will
attempt to be as self-contained as possible, the subject has become
too large to cover every application of \index{Doppler tomography}Doppler tomography and so I
instead focus upon \index{cataclysmic variable}cataclysmic variable stars. I start this endevour
with a potted history of the development of \index{Doppler tomography}Doppler tomography.

\section{History}
\label{sec:stan}
\begin{figure}
\begin{center}
\includegraphics[width=.5\textwidth]{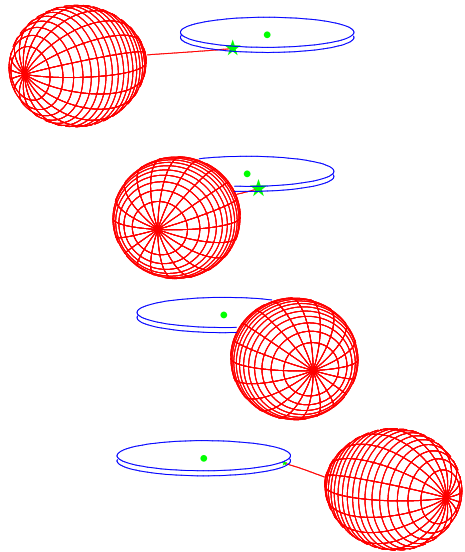}
\end{center}
\caption[]{A schematic illustration of a \index{cataclysmic variable}cataclysmic variable
viewed at four orbital phases.}
\label{CV}
\end{figure}

Although not restricted in application to CVs, since \index{Doppler tomography}Doppler tomography was
developed for CVs and has so far mostly been applied to them (with several
honourable exceptions covered elsewhere in this volume), for completeness
I start by describing our standard picture of CVs. Fig.~\ref{CV} shows a 
schematic representation of our model of non-magnetic CVs with a
\index{white dwarf}white dwarf surrounded by a flat disc, orbiting a tidally-distorted 
main-sequence star. In addition a stream of matter flows from the
main sequence-star and hits the disc in a spot. All of these components,
as well as others, have been seen in Doppler images.
In eclipsing CVs, as seen in Fig~\ref{CV}, the main-sequence star
occults the accreting regions allowing us to locate the chief sources
of emission. It was precisely this that led to the development of the
standard model in the early 1970s [89,72]. In
systems with compact emission sources, the sharp \index{ingress}ingress and \index{egress}egress
features during eclipse allow the source location to be pin-pointed,
and thus, for instance, it is possible to determine the binary mass
ratio from the observed position of the \index{gas stream}gas stream impact
[73,94].  What about more diffuse emission -- the
disc for instance? An obvious possibility is to fit a parameterised
model of the \index{accretion disc!emission}disc emission to light-curves.  However this is plagued
by our lack of good \textit{a priori} models.  For instance, a
symmetric distribution following a power-law in radius is a simple and
obvious choice for modelling discs, but are discs really symmetric and
what if they do not follow power-laws? These problems were solved by
Keith Horne who during his thesis work introduced the powerful idea of
regularised fitting to the study of CVs [39]. Horne
modelled the accretion regions with a grid of many independent pixels,
effectively giving a model of great flexibility. To escape the
degeneracy engendered by such an approach, Horne selected the image of
``maximum \index{entropy}entropy'' where the \index{entropy}entropy $S$ is given (in the simplest
case) by
\begin{equation}
S = - \sum_{i=1}^M p_i \ln p_i .\label{entropy}
\end{equation}
Here $p_i$ is given by 
\begin{equation}
p_i = \frac{I_i}{\sum_{j=1}^M I_j},
\end{equation}
where $I_i$ is the image value assigned to pixel $i$. This is the
method of \index{eclipse map}\index{eclipse mapping}eclipse mapping.

In \index{eclipse map}\index{eclipse mapping}eclipse mapping a one-dimensional light curve leads to 
two-dimensional map. Emission lines on the other hand, as I will
explain, contain the extra dimension of velocity and so can give
better constrained images (albeit with some disadvantages). Horne 
realised that the formation of the line profile from a disc was 
analogous to the formation of medical \index{X-ray emission}X-rays used in computed 
tomography to image the human brain. This led to the development of 
\index{Doppler tomography}Doppler tomography [48], implemented in the first 
instance in a way analogous to \index{eclipse map}\index{eclipse mapping}eclipse mapping, using maximum \index{entropy}entropy
regularisation. 

\begin{figure}
\begin{center}
\includegraphics[width=.5\textwidth,angle=270]{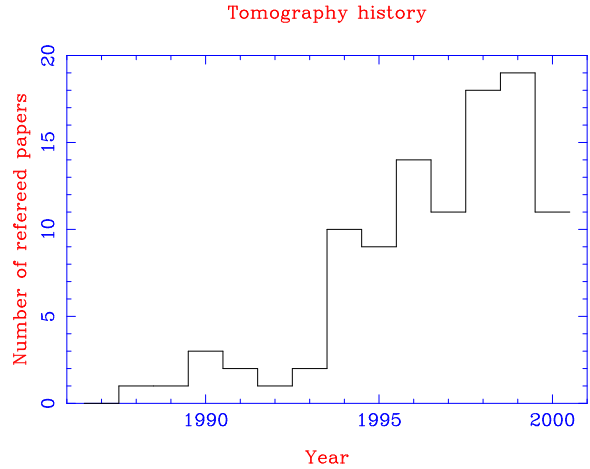}
\end{center}
\caption[]{Numbers of refereed publications using \index{Doppler tomography}Doppler tomography 
versus year of publication.}
\label{history}
\end{figure}

Since that time, and following articles focussing upon
the more accessible \index{filtering}filtered \index{back projection}back-projection inversion 
[40,62], the use of \index{Doppler tomography}Doppler tomography 
has exploded, with over 100 refereed publications making use of it
or containing theoretical simulations of Doppler images
(Fig.~\ref{history}).
\index{Doppler map}Doppler maps have been published for some 16 \index{cataclysmic variable!dwarf nova}dwarf novae, 13 AM~Her stars,
11 DQ~Her stars, 16 \index{cataclysmic variable!nova-like}nova-like variables and 5 black-hole systems, along
with other types such as \index{Algol binaries}Algols and Super-Soft sources. Doppler
tomography is approaching the status of a standard tool; I now begin
discussion of its underlying principles.

\section{The principles of \index{Doppler tomography}Doppler tomography}
\index{Doppler tomography}Doppler tomography arose from a desire to interpret the emission-line
profile variations of accretion discs. Past papers on tomography have
started by describing the formation of the double-peaked profiles from
discs [48]. However, the existence of a disc is not
necessary for \index{Doppler tomography}Doppler tomography. For instance, one of the most
spectacular applications of tomography has been to the AM~Her stars or
\index{cataclysmic variable!polar}polars [67], systems in which the \index{white dwarf}white dwarf has
such a strong \index{magnetic field}magnetic field that there is no disc. Therefore, in this
instance, I will attempt to describe tomography from a more general
perspective.

The key to tomography is first to consider a point-like source of
emission in a binary.  Assuming this has a motion parallel to the
orbital plane of the binary, line emission from such a source will
trace out a sinusoid around the mean velocity of the system. Assuming
one observed such a sinusoid, one could associate it with a particular
velocity vector in the binary, depending upon its phase and
amplitude. The trick of tomography is to cope with any number of such
sinusoids, even when they are so overlapped and blended that one
cannot distinguish one from another. To understand how this is
possible, it is helpful to think of profile formation as a projection
in the mathematical sense of integrating over one dimension of an
$N$-dimensional space to produce an $N-1$ dimensional space.

\subsection{Profile formation by projection}
A given point in the binary can be defined by its spatial position,
but, more usefully in this case, also by its velocity
$(V_x$,$V_y)$. One must be a little careful to define this velocity
which is relative to the \emph{inertial} rather than rotating
frame. Inertial frame velocities are always changing as the binary
rotates, and so in order to define unique values of $V_x$ and $V_y$,
they are measured at a particular orbital phase.  Conventionally this
is taken to be when the inertial frame lines up with the rotating
frame. In the case of CVs it is usual to define the $x$-axis (in the
rotating frame) to point from the \index{white dwarf}white dwarf to the mass donor, and
the $y$-axis to point in the direction of motion of the mass
donor. With this convention, and defining orbital phase zero to be the
moment when the \index{\index{secondary star}secondary star}donor star is closest to us, the \index{radial velocity}radial velocity of
the point in question at orbital phase $\phi$ is
\begin{equation}
V_R = \gamma - V_x\cos 2\pi\phi + V_y\sin 2\pi\phi,
\label{radvel}
\end{equation}
where $\gamma$ is the mean or \index{systemic velocity}systemic velocity of the star. The use of
a single value of $\gamma$ is equivalent to assuming that all motion
is parallel to the orbital plane as mentioned above. 

With these definitions, an ``image'' of the system can be defined as the
strength of emission as a function of velocity, $I(V_x$,$V_y)$. That is,
the flux observed from the system that comes from the velocity element
bounded by  $V_x$ to $V_x +\,dV_x$, $V_y$ to $V_y +\,dV_y$ is given by
$I(V_x$,$V_y)\,dV_x\,dV_y$. I will refer to this as an ``image in
velocity space'' or more simply a ``velocity-space image''. The relation
of this image to the conventional image will be discussed below.

The line flux observed from the system between radial velocities $V$ and
$V+\,dV$ at orbital phase $\phi$ can now be obtained by integration over
all regions of the image that have the correct \index{radial velocity}radial velocity:
\begin{equation}
\int_{-\infty}^\infty \int_{-\infty}^\infty I(V_x,V_y) [g(V - V_R)\,dV]
\,dV_x\,dV_y ,
\end{equation}
where $g$ is a function (of velocity) representing the line profile from 
any point in the image, including instrumental blurring. I assume here
that $g$ is the same at every point, although it is possible to allow it
to vary. The velocity width is divided out to obtain a flux density of 
course, and so the line profile can be expressed as
\begin{equation}
f(V,\phi) = \int_{-\infty}^\infty \int_{-\infty}^\infty I(V_x,V_y) 
g(V - V_R) \,dV_x\,dV_y.
\label{profile}
\end{equation}
Ideally $g$ is narrow, best of all a delta function, thus this equation
picks out all regions of the image close to the line
\[
V = V_R = \gamma - V_x \cos 2\pi\phi + V_y \sin 2\pi\phi.
\]
This is a straight line in $V_x$, $V_y$ coordinates. Different values
of $V$ define a whole family of parallel straight lines across the
image, with a direction dependent upon the orbital phase. With the
definition of velocity and orbital phase described above, orbital
phase $0$ corresponds to a collapse in the positive $V_y$ direction,
phase $0.25$ corresponds to the positive $V_x$ direction, phase $0.5$
to the negative $V_y$ direction etc, with the angle rotating
clockwise.  Thus the formation of the line profile at a particular
phase can be thought of as a projection (or collapse) of the image
along a direction defined by the orbital phase. Note that if this
model is correct, two line profiles taken half-an-orbit apart should
be mirror images of one another. The extent to which this is not the
case is one measure of violations of the basic assumptions made.

\begin{figure}
\begin{center}
\includegraphics[width=.5\textwidth]{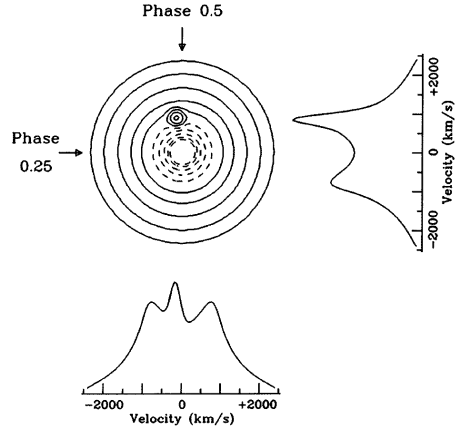}
\end{center}
\caption[]{A model image and the equivalent profiles formed by projection
at angle appropriate to orbital phases $0.25$ (right-most profile) and
$0.5$ (lower profile).}
\label{projection}
\end{figure}
Fig.~\ref{projection} shows a pictorial representation of this process for
two projection angles. The artificial image has been created with a spot 
which can be seen to project into different parts of the profile at 
different phases. Tracing back from the peaks along the projection directions
leads to the position of the original spot. This is in essence how line 
profile information can be used to reconstruct an image of the system.

A series of line profiles at different orbital phases is therefore nothing 
more than a set of projections of the image at different angles. The
inversion of projections to reconstruct the image is known as ``tomography'',
the case of medical X-ray imaging being perhaps the most famous, although
it occurs in many other fields too. I now look at the two methods that have
been applied in the case of \index{Doppler tomography}Doppler tomography.

\subsection{Inversion Methods}
The mathematics of the inversion of projections dates back to the work
of Radon in 1917 [60]. If one knows the function (in my
notation) $f(V,\phi)$ for all $V$ and $\phi$, a linear transformation
-- the \index{Radon transform}Radon transform -- can produce the desired end product,
$I(V_x,V_y)$. In reality, things are not so easy, and we never have
the luxury of knowing the line profiles at all orbital phases,
although one can get close in some cases. With the advent of fast
computers and the development of medical imaging, interest in the
implementation of Radon's transform increased greatly in the 1970s,
and one particular method, that of ``\index{filtering}filtered \index{back projection}back-projection'' found
favour [63].  The original paper on \index{Doppler tomography}Doppler tomography
used an alternative method inherited from \index{eclipse map}\index{eclipse mapping}eclipse mapping, that of
maximum \index{entropy}entropy regularisation.  In this section I describe these two
methods, which are both in use today.  Each has its pros and cons, which I
discuss at the end of the section.

\subsubsection{Filtered \index{back projection}back-projection}
\label{filtbackpro}
The mathematical inversion of Eq.~\ref{profile} is detailed in the Appendix~A.  The
process can be summarized in the following two steps.  First the line
profiles are \index{filtering}filtered in velocity to derive modified profiles,
$\tilde{f}(V,\phi)$. The \index{filtering}filter is applied through a Fourier
transform, multiplication by $|s|/G(s)$, where $G(s)$ is the Fourier 
transform over $V$ of $g(V)$ and $s$ is the frequency in inverse velocity 
units, and finally an inverse \index{Fourier transform}Fourier transform. It can be applied to
one spectrum at a time and is a fairly fast process. 

\begin{figure}
\begin{center}
\includegraphics[width=.35\textwidth]{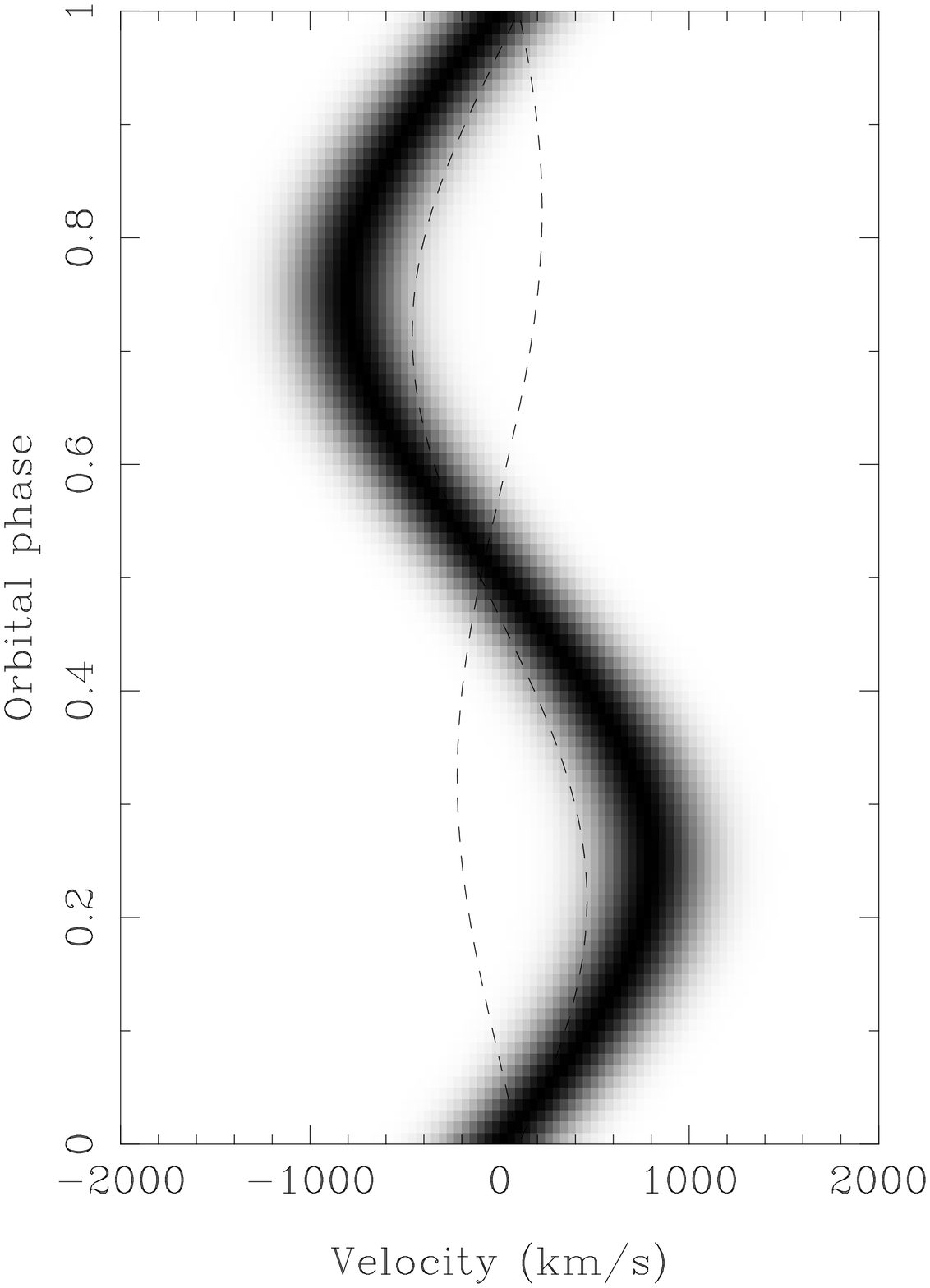}
\hspace{0.06\textwidth}
\raisebox{8mm}{\includegraphics[width=.4\textwidth]{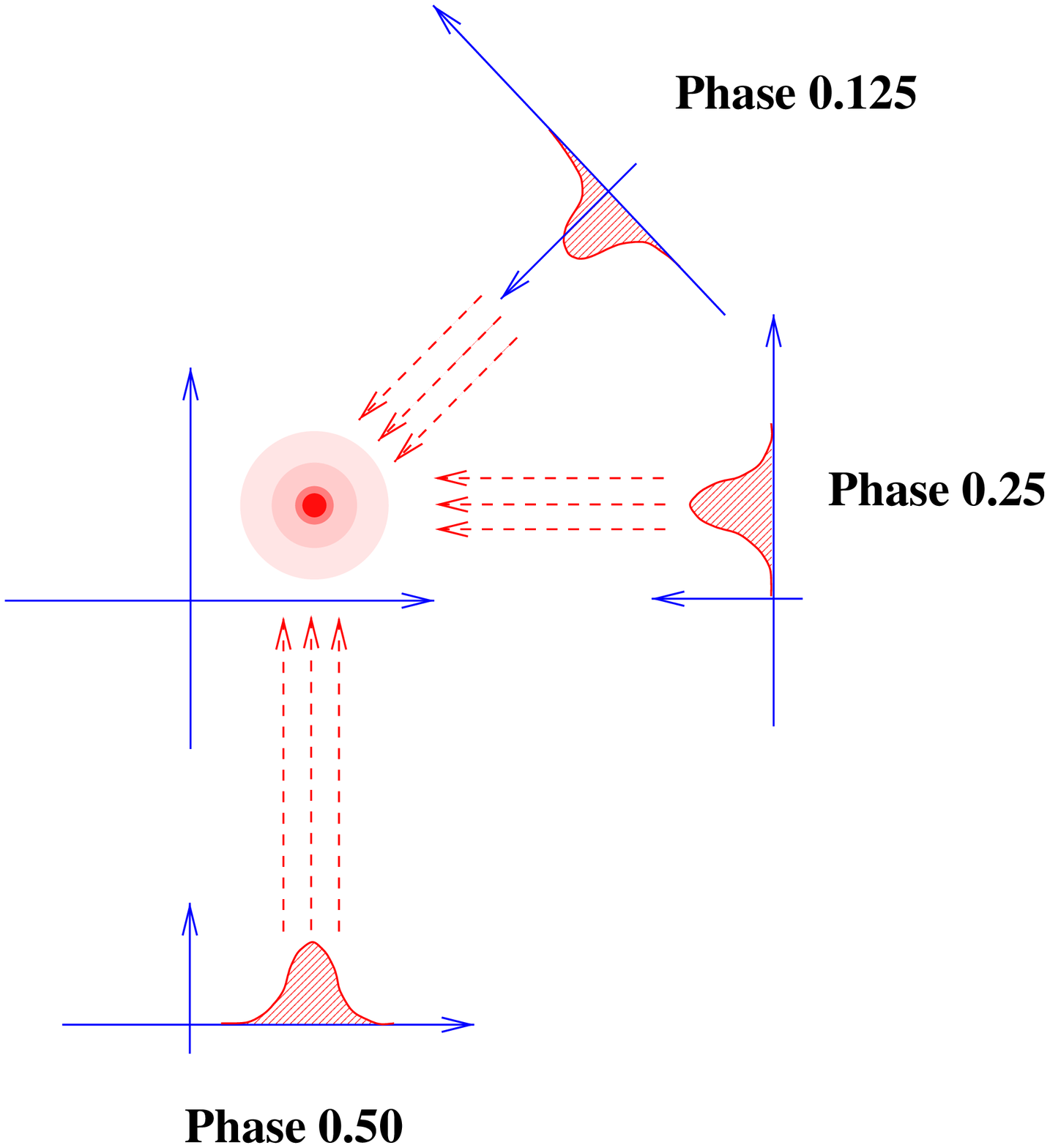}}
\end{center}
\caption[]{Two view of \index{back projection}back-projection: on the left are paths of
integration through trailed spectra (spectra plotted as greyscale images
with time running upwards and wavelength from left to right). A track close to a sinusoidal 
component gives a spot in the final image. On the right three profiles
are smeared back along their original projection directions to give
a spot.}
\label{fig:backpro}
\end{figure}

The second step is that of \emph{\index{back projection}back-projection}:
\begin{equation}
I(V_x,V_y) = \int_0^{0.5} 
\tilde{f}(\gamma-V_x\cos 2\pi\phi+V_y\sin 2\pi\phi,\phi) \,d\phi . 
\label{eq:backpro}
\end{equation}
An intuitive understanding of \index{back projection}back-projection is extremely useful when
trying to make sense of \index{Doppler map}Doppler maps. There are two ways of imagining
the process (see Fig.~\ref{fig:backpro}).  
The first one is perhaps the most obvious from
Eq.~\ref{eq:backpro} which implies that each point in the image can be
built by integration along a sinusoidal path through ``trailed''
spectra (spectra viewed in 2D form with axes of phase versus velocity). 
The particular sinusoid is exactly that which a spot at the
particular place in the image would produce in the trailed
spectrum. This view is illustrated on the left of Fig.~\ref{fig:backpro}.

However, \index{back projection}back-projection is named for another, perhaps more useful,
way of regarding this operation shown on the right of
Fig.~\ref{fig:backpro}.  In effect, Eq.~\ref{eq:backpro} means that the
image is built up by smearing each \index{filtering}filtered profile along the same
direction as the original projection which formed it. This way of
looking at \index{back projection}back-projection shows very clearly why small numbers of
spectra cause linear artifacts in \index{Doppler map}Doppler maps, and one should always
be wary of such features. Similarly, any anomalies, such as unmasked
cosmic rays, dead pixels, \index{flare}flares or unmasked eclipses are liable to
cause streaks across \index{Doppler map}Doppler maps.

If the local line profile $g(V)$ is gaussian then so too is $G(s)$, dropping
to zero at large $s$. Thus the \index{filtering}filter $|s|/G(s)$ will strongly amplify 
high frequencies, and the image will be corrupted by noise. This may be 
familiar when it is realised that division by $G(s)$ is just the standard (and
noise-sensitive) Fourier deconvolution; the presence of $|s|$ in this
case only exacerbates the problem. One can just remove the $G(s)$ term and thus
make no attempt to de-convolve the image. Typically one goes further still
and the \index{filtering}filter applied is $|s| W(s)$, where $W(s)$ is a (typically gaussian) 
``window'' function to cut off high frequencies and therefore limit the 
propagation of noise into the final image. The penalty for this is that the 
final image is a blurred version of the true image. A similar trade-off will
become apparent in the maximum \index{entropy}entropy inversion which I turn to now.

\subsubsection{Maximum Entropy Inversion}
In the original paper presenting \index{Doppler tomography}Doppler tomography, the focus was upon
an alternative to \index{filtering}filtered \index{back projection}back-projection using maximum \index{entropy}entropy 
regularisation [48]. This stemmed in part from the earlier 
development of \index{eclipse map}\index{eclipse mapping}eclipse mapping but also because I originally developed 
\index{Doppler tomography}Doppler tomography in spatial coordinates in which the projection becomes
one over a set of curves rather than straight lines [47].
However, velocity space is nearly universally used now, and the linear
inversion has become more commonly used on the whole. Is another inversion 
method useful? I think the answer is yes, and will discuss why in detail
below. First of all, I describe the maximum \index{entropy}entropy method in some more
detail.

The application of maximum \index{entropy}entropy to \index{Doppler tomography}Doppler tomography is very similar
to the \index{eclipse map}\index{eclipse mapping}eclipse mapping case: a grid of pixels spanning velocity space is
adjusted to achieve a target goodness-of-fit, measured by $\chi^2$. In
general there are an infinite number of such images and so the image
of maximum \index{entropy}entropy is selected. A refined form of \index{entropy}entropy which measures
departures from a ``default'' image [39] is used:
\begin{equation}
S = - \sum_{i=1}^M p_i \ln \frac{p_i}{q_i} .
\end{equation}
Here all symbols are as before with the addition of
\begin{equation}
q_i = \frac{D_i}{\sum_{j=1}^M D_j},
\end{equation}
where $\mathbf{D}$ is the default image. The default image is important
in \index{eclipse map}\index{eclipse mapping}eclipse mapping (see Baptista, this volume) but less so in Doppler
tomography. Usually a moving default is used, computed as a blurred version
of the image. This constrains the map to be smooth on scales shorter
than the blurring, but fixed by the data alone on larger scales. For
reasonable data, the choice of default appears to have little effect.
Indeed, it would be my guess that neither does the form of $S$, and
that its most important role in this case is to allow a unique
solution to be found.

\subsubsection{Relative merits of the two inversions}
Table~\ref{tab:memfbp} compares the Maximum Entropy Method (\index{maximum entropy}MEM)
and Filtered Back-Projection (FBP) inversion methods;
plus and minus signs indicate pros and cons respectively.
\begin{table}
\begin{center}
\begin{tabular}{lll}
Characteristic        & \index{maximum entropy}MEM              & FBP \\
\hline
Controlling parameter      & $\chi^2$     & FWHM of noise \index{filtering}filter\\ 
Processor time             & $-$ ($22$ sec)$^\dagger$ & $+$ ($0.7$ sec) \\
Comparison with data       & $++$         & $--$ \\
Consistency of noise level & $-^\ddagger$ & $+$ \\
Flexibility                & $++$         & $--$ \\
\hline
\end{tabular}\\
{\small $^\dagger$100 by 100 image, 100 by 50 data, 300 MHz Pentium II.}
\end{center}

\caption{Comparison of maximum \index{entropy}entropy and \index{filtering}filtered \index{back projection}back-projection
for \index{Doppler tomography}Doppler tomography.} 
\label{tab:memfbp}
\end{table}
First of all, $\chi^2$ in the maximum \index{entropy}entropy method (\index{maximum entropy}MEM)
is identified as the equivalent of the window \index{filtering}filter in the
\index{filtering}filtered \index{back projection}back-projection (FBP). For instance a low $\chi^2$ forces
the image to become highly structured in order to fit the data better.
Conversely, a high $\chi^2$ allows the image to become smooth and blurred.
Next I compare processing time, which is perhaps the major disadvantage 
of \index{maximum entropy}MEM. In the example given, \index{maximum entropy}MEM took 30 times longer than FBP. Even so, 
for single images the absolute amount of time taken is not large, 
especially when compared with the steps taken to get the data
in the first place, although it can become more significant when trying to
estimate noise, as I will discuss later. Nevertheless, I still regard 
it as a relatively minor disadvantage, and award a single minus. The
next entry ``Comparison with data'' refers to the central role that
$\chi^2$ plays in the \index{maximum entropy}MEM reconstructions which allows one to compare the
predicted data directly to observations. Filtered \index{back projection}back-projection does not
try to achieve a good fit to the data, leaving one uncertain as to how
much better the fit could have been; I regard this as a significant
disadvantage of the method.

Propagation of noise into \index{maximum entropy}MEM images can be problematical when one is
comparing, say multiple datasets. The reason is that if one sets the
same $\chi^2$ level for each inversion, in one case this might be easy
to reach and a rather smooth looking map is the result, while in the next
may have a hard time reaching the desired level at all, resulting in a noisy
map. FBP on the other hand seems to give an images of similar appearance
for the same window \index{filtering}filter. It was for this reason that Marsh \& Duck
[52] used FBP in their tomography of the DQ~Her star, FO~Aqr.
It is possible that this problem could be fixed by iterating towards a
fixed \index{entropy}entropy and minimising $\chi^2$. While this may seem a bit \textit{ad hoc}, 
it may provide a more accurate representation of how one actually sets
$\chi^2$ in practice: it is rather rare to achieve statistically 
acceptable values of $\chi^2$ for data of reasonably good signal-to-noise 
ratio, and often $\chi^2$ is set so that the image is neither too smooth
nor too corrupted by noise.

Flexibility, the final entry in Table~\ref{tab:memfbp}, is another major 
plus point of \index{maximum entropy}MEM. For instance, it is very easy to adapt it to the common
case of blended lines [48,53]. Steeghs (this
volume) presents a nice extension of \index{maximum entropy}MEM to allow for variation of flux
with orbital phase. Another more minor example of this is that it is
easy in \index{maximum entropy}MEM to mask out bad data without the need to interpolate.

For straightforward cases, I think that there is relatively little to
choose between the two methods, although the speed of the \index{filtering}filtered
\index{back projection}back-projection may be advantageous when many maps are being computed.
\index{maximum entropy}MEM has the edge in difficult cases where modifications
of the standard model are needed.

\subsection{Noise in Doppler images}
In the penultimate part of this section, I look at the propagation of
noise into \index{Doppler map}Doppler maps. To an extent Doppler images carry their own
uncertainty estimates in the degree of fluctuation that one sees in
the background, and perhaps this has motivated the lack of a more
rigorous treatment to date. Moreover, it is often the case, as I
remarked above, that \index{Doppler tomography}Doppler tomography cannot achieve a good fit to
the data, and one must assume that systematic errors are
dominant. However, there is still a need to understand noise, with a
common case being the question of the reality of a certain
feature. Appealing to the level of background noise is not always good
enough -- for example, any map will have a highest point, but how is
one to judge whether it is significant?

It has been said that noise cannot be propagated into \index{Doppler map}Doppler maps
[84]. It is in fact straightforward to do so. However,
the important point to understand is that noise in Doppler images is
\emph{correlated}. In more detail it is correlated on short scales but
less so on large scales. This correlation means for example that
single pixel variances are more-or-less useless in defining the amount
of noise on a Doppler image. The correlation is positive on short
scales and so \index{Doppler map}Doppler maps are effectively noisier than an
uncorrelated image with the same variance per pixel; the difference is
significant.

\begin{figure}
\begin{center}
\includegraphics[width=.5\textwidth,angle=270]{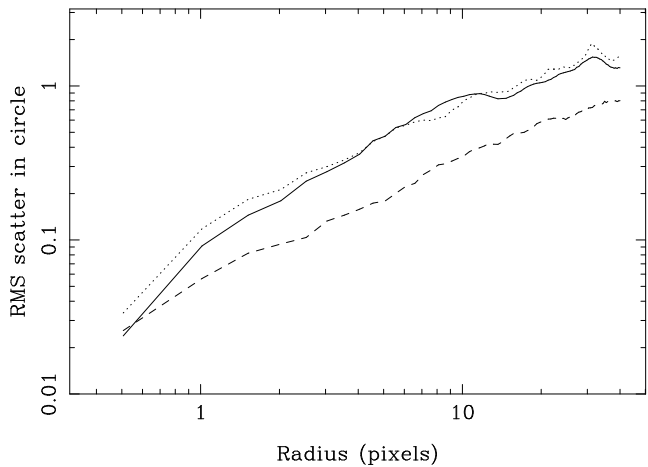}
\end{center}
\caption[]{The figure shows the RMS scatter in simulated Doppler
images measured in circular apertures, as a function of the aperture
radius for three sets of simulations. The solid line shows results
of \index{maximum entropy}MEM inversions with $\chi^2 = 1.0$; the dotted line shows the
FBP results with a \index{filtering}filter of FWHM $=1.0$ in terms of the Nyquist
frequency; the dashed line shows what the result would have been if
the noise were uncorrelated with the same variance per pixel as the
\index{maximum entropy}MEM simulations.}
\label{fig:scatter}
\end{figure}

One of the best ways to appreciate the correlation is to view
animations running simulated images in movie form. Since this is
not possible in print, I illustrate the consequence by plotting
the scatter in circular apertures in Fig.~\ref{fig:scatter}
In this figure, although the RMS scatter of the uncorrelated images
was adjusted to match the \index{maximum entropy}MEM images pixel-by-pixel, the scatter
at first grows more quickly with radius in the FBP and \index{maximum entropy}MEM images,
leading to an RMS about 3 times larger at large radii. This illustrates
the positive correlation at small separations. Ignoring this correlation
would lead to a very significant underestimate of the true noise.
In this case for radii larger than $\approx 4$ pixels, all three lines 
are roughly parallel, demonstrating the weak correlation on large scales. 
The final point is the very similar behaviour of the FBP and \index{maximum entropy}MEM plots:
the two methods lead to a very similar propagation of noise into the
image.

\begin{figure}
\begin{center}
\includegraphics[width=.5\textwidth,angle=270]{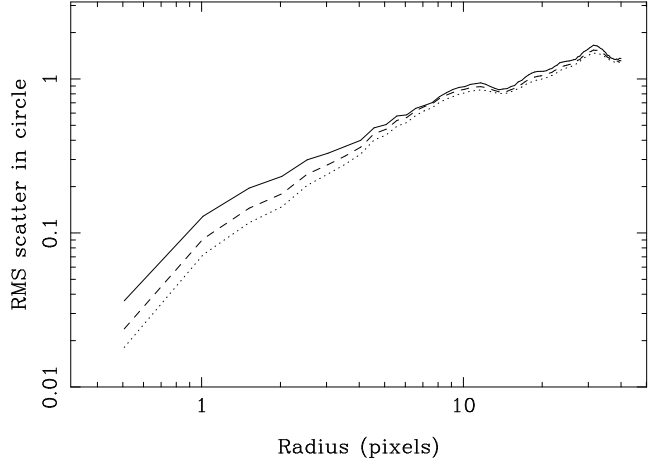}
\end{center}
\caption[]{The figure shows the RMS scatter in \index{maximum entropy}MEM simulations
measured in circular apertures for $\chi^2 = 0.9$ (solid line),
$1.0$ (dashed), and $1.1$ (dotted).}
\label{fig:scatter_chi}
\end{figure}

The exact pattern of noise is dependent upon the controlling
parameter, $\chi^2$ or the \index{filtering}filter FWHM. While driving down $\chi^2$
may force a better fit and higher resolution, a penalty is paid in
increased noise.  Fig.~\ref{fig:scatter_chi} shows this effect. In
this case as $\chi^2$ is lowered from $1.1$ to $0.9$, the noise at the
smallest scales increases by a factor of two; large scales are almost
unaffected. Very similar behaviour is seen for different \index{filtering}filter widths
for \index{filtering}filtered \index{back projection}back-projection.

Figures~\ref{fig:scatter} and \ref{fig:scatter_chi} were made by
adding noise to a simulated dataset and then reconstructing images.  A
similar method was employed by Hessman \& Hopp in their analysis of
GD~552 [35]. In practice, the ``\index{bootstrap method}bootstrap'' method
[15] is preferable. In this technique, real data
are used to generate artificial data by randomly selecting, \emph{with
replacement}, $N$ new points from $N$ old points. The beauty of this
method is that it automatically accounts for the true statistics of
the data and one is not adding extra noise. Consider then an image
with a feature one wants to characterise. As long as an explicit
measurement of the feature can be devised, then multiple \index{bootstrap method}bootstrap
runs can be used to generate an uncertainty as well.

\begin{figure}
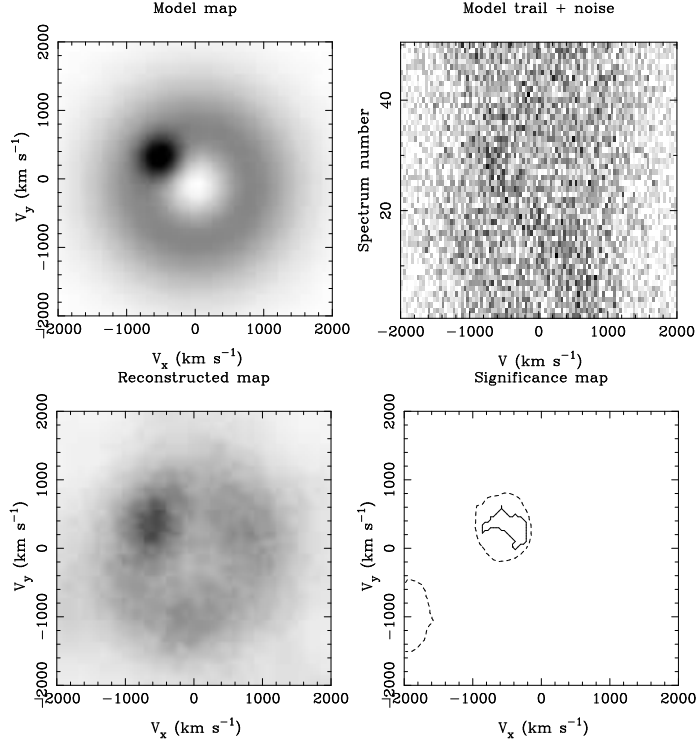

\begin{center}
\includegraphics[width=.4\textwidth,angle=270]{dopp_noise_model.ps}
\includegraphics[width=.4\textwidth,angle=270]{dopp_noise_trail.ps}
\includegraphics[width=.4\textwidth,angle=270]{dopp_noise_map.ps}
\includegraphics[width=.4\textwidth,angle=270]{dopp_noise_signif.ps}
\end{center}
\caption{The panels show an initial test map (upper-left),
the equivalent trailed spectrum with noise added (upper-right), 
the reconstructed image (lower-left) and finally the lower-right
panel shows contours encircling regions of the reconstruction 
above 99\% of the 1000 trials (dashed) and 100\% of them (solid).}
\label{fig:boot}
\end{figure}
As an example of how one might use the \index{bootstrap method}bootstrap method, consider
an image showing evidence for a bright-spot, with some
question as to the significance of the feature (Fig.~\ref{fig:boot}).
First one needs a method of measuring the strength of such a feature.
The method I use here is first to subtract off the symmetric
part of the image and then to measure the flux in a circular aperture
centred on the spot. I then use the following procedure to determine
the significance of the feature:
\begin{enumerate}
\item Generate a large number ($\sim 1000$) of new datasets from the
original data by \index{bootstrap method}bootstrap resampling (a fast procedure).

\item Compute maps for each of these in the same manner as for the 
true map (here the CPU penalty of \index{maximum entropy}MEM is paid in full).

\item Subtract the true map from each boot-strapped map to obtain
difference maps.

\item Measure the flux of each difference map in the same way as before i.e.
by subtracting the symmetric part of each image and computing
the flux in a circular aperture.

\item Finally, rank the observed flux relative to the fluxes measured
from the difference maps.
\end{enumerate}
This procedure generates a set of values showing the stochastic
noise level in the circular aperture. If carried out for apertures
centred on every pixel in the image, one can generate a ``significance
map'' which shows the fraction of fluctuation values that the observed
value exceeds. The result of such a scheme is shown in Fig.~\ref{fig:boot}
where an artificial image shown in the top-left led, after the addition
of gaussian noise, to the trailed spectrum shown in the top-right.
The ``true map'' referred to above is displayed in the lower-left
and shows evidence for a spot in the upper-left quadrant. Carrying
out the \index{bootstrap method}bootstrap computations with 1000 trials leads to the significance
map of the lower-right. This shows that the region of the spot is higher
than all 1000 of the simulated datasets, whereas no other part of the
image is. In this case the circular aperture used has a radius of
$200\,$\kms, and so can be fitted $\sim 100$ times over into the image.
Taking these to be independent, then there is $\sim10$\% chance that
one region will exceed all 1000 trials. More trials would be needed
to establish the reality of the feature more firmly, but the principle
is clear.

Similar measurements are easily imagined, for instance, one could
perhaps fit the position of the spot [35], and
subsequently obtain uncertainty estimates from \index{bootstrap method}bootstrapping. All that
is necessary is that the measurement is precisely defined and applied
in the same way to the true and \index{bootstrap method}bootstrap maps.

\subsection{Axioms of Doppler Tomography}
\index{Doppler tomography}Doppler tomography rests on certain approximations to reality
that are, at best, only partially fulfilled. Violation of these
approximations does not mean that the resulting maps are useless, but
everyone who carries out \index{Doppler tomography}Doppler tomography should be aware of its
limitations. Thus in this final part of this section, I list
the ``axioms'' that underly the method:
\begin{enumerate}
\item All points are equally visible at all times.
\item The flux from any point fixed in the rotating frame is constant. 
\item All motion is parallel to the orbital plane.
\item All velocity vectors rotate with the binary star.
\item The intrinsic width of the profile from any point is negligible.
\end{enumerate}
Exceptions exist to each of these. For instance, emission on the
mass donor violates the first axiom, while \index{outburst}outbursts clearly
contravene (2). \index{Doppler tomography}Doppler tomography is an interpretation of the data
within a specific model of binary systems and only applies inasmuch
as the model itself does.

\section{The application of \index{Doppler tomography}Doppler tomography to CVs}
I now move on to discuss \index{Doppler tomography}Doppler tomography in the context of 
\index{cataclysmic variable}cataclysmic variable stars. I will mainly focus upon results,
but before doing so I consider the interpretation of 
\index{Doppler map}Doppler maps in the case of CVs.

\subsection{Understanding \index{Doppler map}Doppler maps}
\begin{figure}
\begin{center}
\includegraphics[width=.5\textwidth]{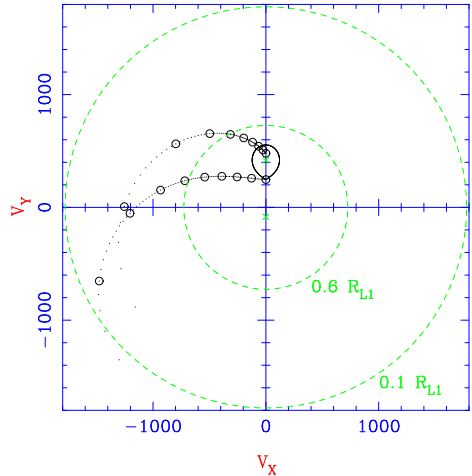}
\end{center}
\caption[]{A schematic of some key components in velocity coordinates.}
\label{velocity_space}
\end{figure}

On the basis of the standard model presented in section~\ref{sec:stan}
one can easily predict the locations of the various components
in velocity-space; Fig.~\ref{velocity_space} shows 
some of the key components
of a CV represented in velocity space. The \index{\index{secondary star}secondary star}donor star is assumed to
co-rotate with the binary, which means that it appears with the same
shape in velocity as it does in position coordinates, although rotated
by $90^\circ$ owing to the relation $\vec{v} = \vec{\Omega} \wedge
\vec{r}$ between velocity and position for ``solid-body''
rotation. This reassuring property is somewhat misleading, since the
disc, which is very definitely not co-rotating with the binary, ends
up being turned inside out so that the inner disc is at large
velocities while the outer disc appears as a ring at low velocity. The
\index{gas stream}gas stream is plotted twice: once with its true velocity and once with
the velocity of the disc along its path; one can also imagine
intermediate cases. The positions of all these components is fully
specified if the projected orbital velocities of the two stars, $K_1$
and $K_2$, and the orbital phase are known. The overall scale is set
by $K_1+K_2$; their ratio, which is the \index{mass ratio}mass ratio $q = K_1/K_2 =
M_2/M_1$, defines the detailed shape of the stream and \index{Roche equipotential}Roche lobe. The
orbital phase sets the orientation of the image, and if it is not
known the image will be rotated by an unknown amount relative to the
``standard'' orientation shown in Fig.~\ref{velocity_space}.
This is not uncommon: for instance if the orbital phase is based upon
emission line measurements, it is typically delayed by $0.05$ -- $0.1$
cycles with respect to the true ephemeris. This causes an anti-clockwise
rotation of the image by an equivalent number of turns. The published
map of LY~Hya [78] is a nice example of this phenomenon.

Although velocity coordinates simplify the picture of line profile
formation, it is simple enough to invert into position coordinates --
indeed this is how I originally computed Doppler images
[47]. All that is required is a specification of the
velocity at every point in the system. However, I abandoned this
approach for two reasons. First, the translation between velocity and
position is often not known. In fact, perhaps it is never known, given
that it is likely that deviations from keplerian flow occur. This
means that position maps would require recomputation each time system
parameters were updated. Second, the same place in the system can
produce emission at more than one velocity. This is not an abstract
possibility, but happens in almost every system that has been
imaged. There are many examples of bright-spot emission from the gas
stream while the disc at the same location produces emission at a
completely different velocity. If such data is imaged into position
coordinates on the basis of keplerian rotation, a spot of emission
would be produced at a spurious location in the disc. Sticking to
velocity coordinates is a reminder of these potential difficulties of
interpretation. Only in eclipsing systems is there potential for
disentangling such effects.

While we cannot translate the data into position space, there is 
no difficulty in translating any theoretical model into velocity 
coordinates. Indeed, ideally, the theory--data comparison should be
made by predicting trailed spectra, doing away with the need for
\index{Doppler map}Doppler maps altogether. However, \index{Doppler map}Doppler maps still have a r\^{o}le
in that theoretical models are not good enough to predict all the
peculiarities of real systems, and comparison is easier in the 
half-way house of velocity space. 

The idea of translating to velocity space also applies to how one
should think about \index{Doppler map}Doppler maps. Rather than trying to translate
features of maps mentally from velocity to position coordinates, one
should try to think of various components and imagine where they would
appear in velocity space. The difference may seem slight, but it is a
significant one. With that said, I now turn to look at some results.

\subsection{Doppler imaging results}
There are now a large number of examples of \index{Doppler tomography}Doppler tomography,
covering CVs along with other types of binary as well, such as \index{Algol binaries}Algols
and X-ray transients, the latter being very similar to CVs in many
ways [51,8]. Rather than spend space
covering these in detail when the original references do so already,
as do other contributions in this volume, I have decided to devote
this section mainly to highlights based upon a literature survey of as
many published Doppler images as I could find. The results of this
survey are tabulated in Tables~\ref{tab:part1}, \ref{tab:part2} and
\ref{tab:part3} contained in Appendix~B where I list systems with
published maps, which lines were mapped, the spectral resolution, the
\index{outburst}outburst state and some indication of the appearance of the maps.

\subsubsection{Spiral shocks}
The discovery by Steeghs et al. [77] of \index{spiral arms}spiral shocks
in the \index{cataclysmic variable!dwarf nova}dwarf nova IP~Peg is perhaps the most significant
result from \index{Doppler tomography}Doppler tomography applied to CVs.
\begin{figure}
\begin{center}
\includegraphics[width=.7\textwidth,bbllx=40,bblly=196,bburx=535,bbury=438,clip=]{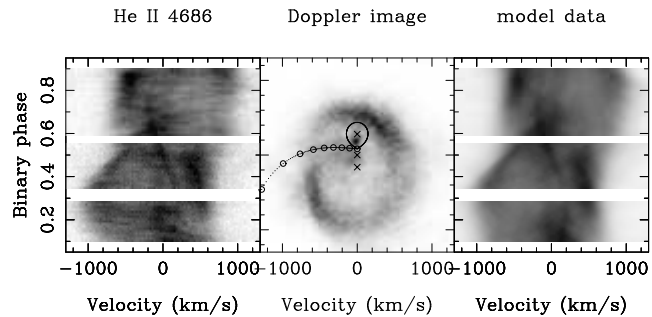}
\end{center}
\caption[]{Spiral shocks in IP~Peg [27].}
\label{fig:spiral}
\end{figure}
These shocks appear to
be present in all \index{outburst}outbursts of IP~Peg, including (with hindsight!)
pre-discovery \index{outburst}outburst data [50]; Fig.~\ref{fig:spiral} shows
one example. There is corroborating
evidence from other systems such as SS~Cyg [76],
V347~Pup [82] and EX~Dra [42], although none of these are as convincing 
as IP~Peg. 

\subsubsection{Stream emission in \index{cataclysmic variable!polar}polars}
Beautiful work by Schwope and others
[66,67,30,68,71]
has revealed the \index{gas stream}gas stream in the polar class of cataclysmic
variables in which the \index{white dwarf}white dwarf is magnetically locked to the mass
\index{\index{secondary star}secondary star}donor star. Although initially ballistic, there is evidence for the
influence of the field upon the stream and some emission from the gas
\begin{figure}
\begin{center}
\includegraphics[width=.6\textwidth,angle=270]{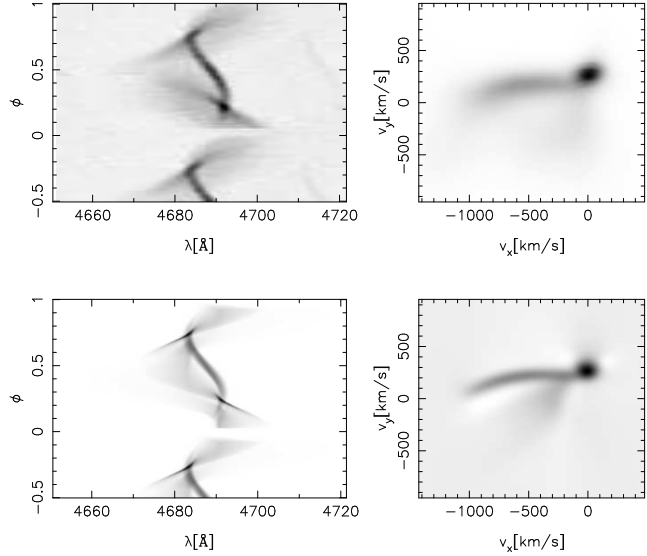}
\end{center}
\caption[]{Stream-stripping in HU~Aqr [30]. The upper panels
show the HeII 4686 line and \index{Doppler map}Doppler map; the lower panels show simulated data
based upon a simple model in which gas is pulled of the stream and threaded
onto the \index{magnetic field}magnetic field.}
\label{fig:huaqr}
\end{figure}
as it hurtles down towards the \index{white dwarf}white dwarf (see Fig.~\ref{fig:huaqr}). 
Such work has tremendous scope for teaching us about the stream/magnetosphere
interaction. Changes in the appearance of maps between high and low
states have been seen in HU~Aqr [67] and further
observation of differing states should tell us how the plasma/field
interaction varies with \index{accretion rate}accretion rate.

\subsubsection{Bright-spots}
\begin{figure}
\begin{center}
\includegraphics[width=.5\textwidth,angle=270]{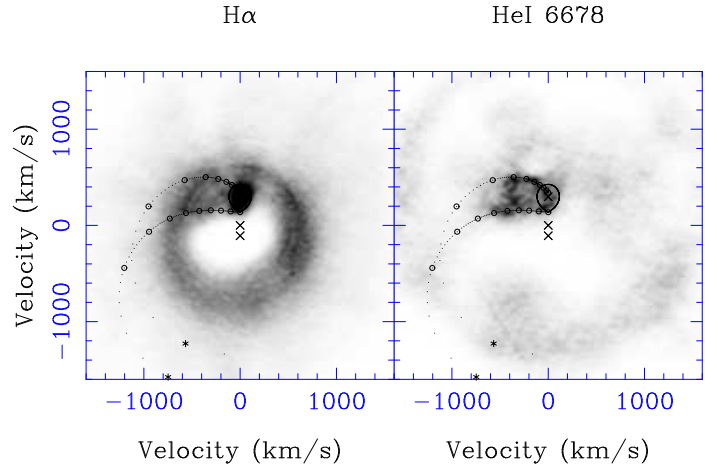}
\end{center}
\caption[]{The peculiar bright-spot of U~Gem observed in January 2000 (Unda et al in prep).}
\label{fig:ugem}
\end{figure}

Many systems show bright-spots in \index{Doppler map}Doppler maps, classic cases being
WZ~Sge [75] and GP~Com [53]. The
locations of the spots are interesting. In some cases they line up
with the stream's velocity [50], while others are
closer to the disc's velocity [93]. Still others adopt a
position half-way between the disc and stream velocities
[49,75] (see also Fig~\ref{fig:ugem}). 
There is some evidence for extended
cooling following the spot [75] and there is potential
for examining this with lines of differing excitation.

\subsubsection{Emission from the \index{secondary star}secondary star}
Many \index{Doppler map}Doppler maps show emission from the \index{secondary star}secondary star 
[50,49,23,57,66]. 
Some of this may be related to magnetic activity of these rapidly rotating stars,
but there is little doubt that much of it is caused by \index{irradiation}irradiation by
the inner disc. The disc systems are interesting in this respect since one
can expect the disc to cast a shadow over the equator of the mass donor.
This seems indeed to be the case [28,57] and perhaps has the
potential to tell us about the vertical structure of the disc.

\subsubsection{Missing discs in novalikes}
Nova-like variable stars are thought to be classic examples of
steady-state systems. Spectroscopically however they have proved
hard to understand. A particular peculiarity of these stars is
that it is often difficult to see any sign of a disc in these
systems. Instead the \index{Doppler map}Doppler maps are often dominated by a
single structure-less blob. This blob tends to be located in the
lower-left quadrant of the map for the Balmer and HeI lines, but
rather closer to the expected location of the \index{white dwarf}white dwarf for
HeII. Several explanations for these phenomena have been proposed
but none of them are compelling in my view.

\section{The future of \index{Doppler tomography}Doppler tomography}
The future development of \index{Doppler tomography}Doppler tomography splits into extensions of
the method and the acquisition of further datasets. Dealing first of
all with the latter, it is evident from Tables~\ref{tab:part1},
\ref{tab:part2} and \ref{tab:part3} that there are several areas where
improvements are possible. For instance more lines could be covered,
especially metal lines (e.g. CaII), \index{ultraviolet}ultraviolet and infra-red
lines. H$\alpha$ has received relatively little attention, but when it
has been looked at, often appears peculiar [81,76]. 
In the case of \index{cataclysmic variable!dwarf nova}dwarf novae, the discovery of \index{spiral arms}spiral shocks makes extended coverage of \index{outburst}outbursts of
considerable interest. 

Multi-epoch tomography is probably the most gaping hole because it is
hard to make a coherent picture of the many one-off maps that have
been published to date. The resolution of maps published to date is
poor or moderate in many cases, limiting their usefulness. The
ultimate limit is set by thermal broadening, but we are very far off
this in most cases, especially for heavy elements. Pushing to high
resolution is not trivial because of the concomitant need to shorten
the exposure time owing to smearing, but it is thoroughly feasible. 
Exposures of length $t$ can be thought of as blurring the image by a rotation of
\[ 360^\circ \frac{t}{P} \]
where $P$ is the orbital period. For a feature a speed $K$ from the
centre of mass, this will match the spectral resolution $\Delta V$ when
\[ \frac{t}{P} \sim \frac{1}{2\pi} \frac{\Delta V}{K} .\]
Thus if $\Delta V$ is lowered by a factor of two, the exposure time must
also be reduced by a factor two. As a concrete example, consider trying
to image the mass donor in a system where $K_2 = 400\,$\kms. Typical equatorial
velocities in CVs are $\approx 100\,$\kms, so we may attempt to obtain data
with $\Delta V = 10\,$\kms. We would then need $\Delta \phi = t/P < 0.004$,
equivalent to $t < 30 \,\mathrm{s}$ for a system of $P = 2$ hours. For many
CVs, this will require 8m-class telescopes.

Various extensions are possible, and have already been developed in
some cases. Standard \index{Doppler tomography}Doppler tomography does not treat the geometry of
the \index{\index{secondary star}secondary star}donor star correctly. It is straightforward to fix this
[64].  Bobinger et al.\ [3]
describe a method for simultaneous Doppler and \index{eclipse map}\index{eclipse mapping}eclipse mapping of
emission lines in which a single image is computed to fit both spectra
and light curves of the lines, with a keplerian velocity field used to
translate between position and velocity space. It is difficult to
evaluate whether the spectra or fluxes dominate the final maps, but it
is clear that spectral information does alleviate the degenerate
nature of \index{eclipse map}\index{eclipse mapping}eclipse mapping. Of course, the need to assume a particular
velocity field is a disadvantage.  An attempt has been made to avoid
this by simultaneously adjusting a spatial image and a
position--velocity map to fit spectra of eclipsing systems
[1].  In this method, spectra out of eclipse
serve to fix the velocity space image as usual, which is then
translated to position space through the eclipse information. The
technique was able to recover a $V \propto R^{-1/2}$ relation from
spectra of V2051~Oph, but as developed it could not handle the
difficult case of the same place producing emission at more than one
velocity. Finally Steeghs (this volume) describes a new modification
which allows orbital \index{variability}variability to be included in Doppler images.
As this is so common, it has considerable potential.

\section{Conclusions}
I have reviewed the principles and practice of the analysis method
of \index{Doppler tomography}Doppler tomography which helps the interpretation of the complex
line profile variations from close binary stars. The key discoveries from
the application of this technique are the \index{spiral arms}spiral shocks in \index{outburst}outbursting
\index{cataclysmic variable!dwarf nova}dwarf novae and the stream/\index{magnetic field}magnetic field interaction in the polar class
of \index{cataclysmic variable}cataclysmic variable stars, but \index{Doppler tomography}Doppler tomography has taught us much
about the stream/disk interaction and \index{irradiation}irradiation of the \index{\index{secondary star}secondary star}donor star too.

For the future, efforts need to be made to acquire multi-epoch datasets
for tomography as these are wholly lacking at present. Following that
higher spectral and temporal resolution data are needed to exploit
tomography to its limit.

\section*{References}
\addcontentsline{toc}{section}{References}

\begin{itemize}
\item[1.] Billington, I., 1995, PhD Thesis, Oxford University.

\item[2.] Billington, I., Marsh, T.R., Dhillon, V.S., 1996, \mnras, 278, 673--682.

\item[3.] Bobinger, A.  et al., 1999, \aap, 348, 145--153.

\item[4.] Burwitz, V. et al., 1998, \aap, 331, 262--270.

\item[5.] Casares, J. et al., 1995, \mnras, 274, 565--571.

\item[6.] Casares, J., Charles, P.A., Marsh, T.R., 1995, \mnras, 277, L45--L50.

\item[7.] Casares, J. et al., 1996, \mnras, 278, 219--235.

\item[8.] Casares, J. et al., 1997, New Astronomy, 1, 299--310.

\item[9.] Casares, J. et al., 1997, New Astronomy, 1, 299--310.

\item[10.] Catal\'an, M.\ S., Schwope, A.\ D., Smith, R.\ C., 1999, \mnras, 310, 123--145.

\item[11.] Dhillon, V.S., Marsh, T.R., Jones, D.H.P., 1991, \mnras, 252, 342--356.

\item[12.] Dhillon, V.S., Jones, D.H.P., Marsh, T.R., Smith, R.C., 1992, \mnras, 258, 225--240.

\item[13.] Dhillon, V.S., Jones, D.H.P., Marsh, T.R., 1994, \mnras, 266, 859.

\item[14.] Dhillon, V.S., Marsh, T.R., Jones, D.H.P., 1997, \mnras, 291, 694.

\item[15.] Diaconis, P., Efron, B., 1983, Sci.\ Am.\, 248, 96--.

\item[16.] Diaz, M.\ P., Steiner, J.\ E., 1994, \apj, 425, 252--263.

\item[17.] Diaz, M.\ P., Steiner, J.\ E., 1994, \aap, 283, 508--514.

\item[18.] Diaz, M.\ P., Steiner, J.\ E., 1995, \aj, 110, 1816.

\item[19.] Diaz, M.\ P., Hubeny, I., 1999, \apj, 523, 786--796.

\item[20.] Dickinson, R.\ J. et al., 1997, \mnras, 286, 447--462.

\item[21.] Gaensicke, B.\ T. et al., 1998, \aap, 338, 933--946.

\item[22.] Harlaftis, E.T., Marsh, T.R., Dhillon, V.S., Charles, P.A., 1994, \mnras, 267, 473.

\item[23.] Harlaftis, E.T., Marsh, T.R., 1996, \aap, 308, 97--106.

\item[24.] Harlaftis, E.\ T., Horne, K., Filippenko, A.\ V., 1996, \pasp, 108, 762.

\item[25.] Harlaftis, E.\ T., Steeghs, D., Horne, K., Filippenko, A.\ V., 1997, \aj, 114, 1170--1175.

\item[26.] Harlaftis, E., Collier, S., Horne, K., Filippenko, A.\ V., 1999, \aap, 341, 491--498.

\item[27.] Harlaftis, E. T. et al., 1999, \mnras, 306, 348--352.

\item[28.] Harlaftis, E., 1999, \aap, 346, L73--L75.

\item[29.] Hastings, N.\ C. et al., 1999, \pasp, 111, 177--183.

\item[30.] Heerlein, C., Horne, K., Schwope, A.D., 1999, \mnras, 304, 145--154.

\item[31.] Hellier, C., Robinson, E.L., 1994, \apjl, 431, L107--L110.

\item[32.] Hellier, C., 1996, \apj, 471, 949.

\item[33.] Hellier, C., 1997, \mnras, 288, 817--832.

\item[34.] Hellier, C., 1999, \apj, 519, 324--331.

\item[35.] Hessman, F.V., Hopp, U., 1990, \aap, 228, 387--398.

\item[36.] Hoard, D.\ W., Szkody, P., 1996, \apj, 470, 1052.

\item[37.] Hoard, D.\ W., Szkody, P., 1997, \apj, 481, 433.

\item[38.] Hoard, D.\ W. et al., 1998, \mnras, 294, 689.

\item[39.] Horne, K., 1985, \mnras, 213, 129--141.

\item[40.] Horne, K., 1991, Fundamental Properties of Cataclysmic Variable Stars: Proc. 12th North American Workshop on CVs and Low Mass X-Ray Binaries, ed. A. W. Shafter (San Diego: San Diego State Univ.), , 23--.

\item[41.] Howell, S.\ B., Ciardi, D.\ R., Dhillon, V.\ S., Skidmore, W., 2000, \apj, 530, 904--915.

\item[42.] Joergens, V., Spruit, H.\ C., Rutten, R.\ G.\ M., 2000, \aap, 356, L33--L36.

\item[43.] Joergens, V. et al., 2000, \aap, 354, 579--588.

\item[44.] Kaitchuck, R. H. et al., 1994, \apjs, 93, 519--530.

\item[45.] Kaitchuck, R.\ H., Schlegel, E.\ M., White, J.\ C., Mansperger, C.\ S., 1998, \apj, 499, 444.

\item[46.] Littlefair, S.\ P., Dhillon, V.\ S., Howell, S.\ B., Ciardi, D.\ R., 2000, \mnras, 313, 117--128.

\item[47.] Marsh, T.R., 1985, PhD Thesis, Cambridge University.

\item[48.] Marsh, T.R., Horne, K., 1988, \mnras, 235, 269--286.

\item[49.] Marsh, T. R. et al., 1990, \apj, 364, 637--646.

\item[50.] Marsh, T.R., Horne, K., 1990, \apj, 349, 593--607.

\item[51.] Marsh, T.R., Robinson, E.L., Wood, J.H., 1994, \mnras, 266, 137.

\item[52.] Marsh, T.R., Duck, S.R., 1996, New Astronomy, 1, 97--119.

\item[53.] Marsh, T.R., 1999, \mnras, 304, 443--450.

\item[54.] Martell, P.\ J., Horne, K., Price, C.\ M., Gomer, R.\ H., 1995, \apj, 448, 380.

\item[55.] Mennickent, R.\ E., Diaz, M., 1996, \aap, 309, 147--154.

\item[56.] Mennickent, R.\ E., Diaz, M.\ P., Arenas, J., 1999, \aap, 352, 167--176.

\item[57.] Morales-Rueda, L., Marsh, T.R., Billington, I., 2000, \mnras, 313, 454--460.

\item[58.] Nogami, D., Masuda, S., Kato, T., Hirata, R., 1999, \pasj, 51, 115--125.

\item[59.] North, R.\ C. et al., 2000, \mnras, 313, 383--391.

\item[60.] Radon, J., 1917, Ber.\ Verh. S\"{a}chs.\ Akad.\ Wiss. Leipzig Math.\ Phys.\ K1, 69, 262--277.

\item[61.] Ratering, C., Bruch, A., Diaz, M., 1993, \aap, 268, 694--704.

\item[62.] Robinson, P.F.L., Marsh, T.R., Smak, J., 1999, Accretion Disks in Compact Stellar Systems, Edited by J. Craig Wheeler. World Scientific, ISBN 981-02-1273-9 (1993)., , 75--116.

\item[63.] Rowland, S.W., 1979, Image Reconstruction from Projections, , 8--79.

\item[64.] Rutten, R.G.M., Dhillon, V.S., 1994, \aap, 288, 773--781.

\item[65.] Schmidt, G. D. et al., 1999, \apj, 525, 407--419.

\item[66.] Schwope, A.D., Mantel, K.-H., Horne, K., 1997, \aap, 319, 894--908.

\item[67.] Schwope, A. D. et al., 1998, Wild stars in the old west: Proceedings of the 13th North American Workshop on Cataclysmic Variables and Related Objects, eds. Howell, S. Kuulkers, E., Woodward, C., astro-ph/9708228, , 44--59.

\item[68.] Schwope, A. D. et al., 2000, \mnras, 313, 533--546.

\item[69.] Shafter, A.\ W., Veal, J.\ M., Robinson, E.\ L., 1995, \apj, 440, 853.

\item[70.] Shahbaz, T., Wood, J.\ H., 1996, \mnras, 282, 362--372.

\item[71.] Simic, D. et al., 1998, \aap, 329, 115--130.

\item[72.] Smak, J., 1971, Acta Astronomica, 21, 15.

\item[73.] Smak, J., 1979, Acta Astronomica, 29, 309.

\item[74.] Smith, D.A., Dhillon, V.S., Marsh, T.R., 1998, \mnras, 296, 465--482.

\item[75.] Spruit, H.C., Rutten, R.G.M., 1998, \mnras, 299, 768--776.

\item[76.] Steeghs, D., Horne, K., Marsh, T.R., Donati, J.F., 1996, \mnras, 281, 626--636.

\item[77.] Steeghs, D., Harlaftis, E.T., Horne, K., 1997, \mnras, 290, L28--L32.

\item[78.] Still, M.D., Marsh, T.R., Dhillon, V.S., Horne, K., 1994, \mnras, 267, 957.

\item[79.] Still, M.\ D., Dhillon, V.\ S., Jones, D.\ H.\ P., 1995, \mnras, 273, 863--876.

\item[80.] Still, M.\ D., Dhillon, V.\ S., Jones, D.\ H.\ P., 1995, \mnras, 273, 849--862.

\item[81.] Still, M.\ D., 1996, \mnras, 282, 943--952.

\item[82.] Still, M.\ D., Buckley, D.\ A.\ H., Garlick, M.\ A., 1998, \mnras, 299, 545--553.

\item[83.] Still, M.D., Duck, S.R., Marsh, T.R., 1998, \mnras, 299, 759--767.

\item[84.] Still, M.\ D., Steeghs, D., Dhillon, V.\ S., Buckley, D.\ A.\ H., 1999, \mnras, 310, 39--42.

\item[85.] Szkody, P., Armstrong, J., Fried, R., 2000, \pasp, 112, 228--236.

\item[86.] Tovmassian, G.\ H. et al., 1997, \aap, 328, 571--578.

\item[87.] Tovmassian, G.\ H. et al., 1998, \aap, 335, 227--233.

\item[88.] Tovmassian, G.\ H. et al., 2000, \apj, 537, 927--935.

\item[89.] Warner, B., Nather, R.\ E., 1971, \mnras, 152, 219.

\item[90.] Welsh, W.\ F., Horne, K., Gomer, R., 1998, \mnras, 298, 285--302.

\item[91.] White, J.\ C., Honeycutt, R.\ K., Horne, K., 1993, \apj, 412, 278--287.

\item[92.] White, J.\ C., Schlegel, E.\ M., Honeycutt, R.\ K., 1996, \apj, 456, 777.

\item[93.] Wolf, S. et al., 1998, \aap, 332, 984--998.

\item[94.] Wood, J. et al., 1986, \mnras, 219, 629--655.

\end{itemize}

\newpage
\section*{Appendix~A}

In this appendix I show that, as stated in section~\ref{filtbackpro},
Eq.~\ref{profile} can be inverted by application of the \index{filtering}filter
$|s|/G(s)$ followed by the \index{back projection}back-projection of Eq.~\ref{eq:backpro}.
I define the \index{Fourier transform}Fourier transform $F(s)$ of a function $f(x)$, and its
inverse by
\begin{eqnarray*}
 F(s) &=& \int_{-\infty}^\infty f(x) e^{-i 2\pi s x} \,dx .\\
 f(x) &=& \int_{-\infty}^\infty F(s) e^{i 2\pi s x} \,ds .
\end{eqnarray*}
The frequency $s$ here is measured in cycles per unit $x$.
Now take the \index{Fourier transform}Fourier transform over $V$ of the line profile
equation, \ref{profile}:
\begin{eqnarray}
F(s,\phi) &=& \int_{-\infty}^\infty f(V,\phi) e^{-i2\pi s V}\,dV \\
&=&
\int_{-\infty}^\infty \int_{-\infty}^\infty I(V_x,V_y) 
\int_{-\infty}^\infty g(V - V_R) e^{-i2\pi s V}\,dV \,dV_x\,dV_y \\ 
&=& G(s) \int_{-\infty}^\infty \int_{-\infty}^\infty I(V_x,V_y) 
e^{-i2\pi s V_R} \,dV_x\,dV_y.
\end{eqnarray}
Dividing through by $G(s)$, multiplying by $|s|$ and taking the 
inverse \index{Fourier transform}Fourier transform gives the \index{filtering}filtered line profiles
\begin{eqnarray}
\tilde{f}(V,\phi) &=& \int_{-\infty}^\infty \frac{|s| F(s,\phi)}{G(s)} 
e^{i 2\pi s V}\,ds \nonumber \\
&=&  \int_{-\infty}^\infty \int_{-\infty}^\infty I(V_x,V_y) 
\int_{-\infty}^\infty |s| e^{-i2\pi s (V-V_R)}\,ds \,dV_x\,dV_y.
\label{filt}
\end{eqnarray}

Finally, back-project these \index{filtering}filtered profiles according to
Eq.~\ref{eq:backpro}, that is compute the integral
\[ \int_0^{0.5}\tilde{f}(V_R,\phi) \,d\phi ,\]
where
\begin{equation}
V_R = \gamma - V_x\cos 2\pi\phi + V_y\sin 2\pi\phi .
\label{rvel1}
\end{equation}
Putting dashes on various symbols to avoid confusion later, then
the \index{back projection}back-projection integral becomes
\begin{eqnarray}
\int_0^{0.5} \tilde{f}(V_R,\phi) \, d\phi
&=&\int_{-\infty}^\infty \int_{-\infty}^\infty I(V'_x,V'_y) 
\int_0^{0.5} \int_{-\infty}^\infty |s|
e^{-i2\pi s (V_R-V'_R)}\,ds\,d\phi \,dV'_x\,dV'_y \nonumber\\
&=& \int_{-\infty}^\infty \int_{-\infty}^\infty I(V'_x,V'_y) 
\int_0^1 \int_{0}^\infty s
e^{-i2\pi s (V_R-V'_R)}\,ds\,d\phi \,dV'_x\,dV'_y \nonumber\\
&=& \int_{-\infty}^\infty \int_{-\infty}^\infty I(V'_x,V'_y) 
\delta(V'_x-V_x) \delta(V'_y-V_y) \,dV'_x\,dV'_y \nonumber\\
&=& I(V_x,V_y). \label{eq1}
\end{eqnarray} 
The third line above follows from the second after transforming from
polar coordinates $s$ and $\phi$ to cartesian $s_x = s\cos 2\pi\phi$ and
$s_y = s\sin 2\pi\phi$, and using Eqs.~\ref{rvel1} so that
\begin{equation}
\int_0^1 \int_{0}^\infty s
e^{-i2\pi s (V_R-V'_R)}\,ds\,d\phi =
\int_{-\infty}^\infty \int_{-\infty}^\infty
e^{-i2\pi [-(V_x-V'_x)s_x + (V_y-V'_y)s_y]}\,ds_x\,ds_y
\end{equation}
and then the integrals over $s_x$ and $s_y$ separate to give
the two Dirac $\delta$-functions of the penultimate line of Eq.~\ref{eq1}
since
\[ \delta(x) = \int_{-\infty}^\infty e^{\pm i 2\pi s x} \,ds . \]

This justifies the assertions of section~\ref{filtbackpro}.

\newpage
\section*{Appendix~B}

\begin{table}
\caption{\index{Doppler map}Doppler maps of CVs and X-ray novae published in
refereed journals as of September 2000.}
\begin{tabular}{p{0.18\textwidth}p{0.1\textwidth}p{0.11\textwidth}p{0.13\textwidth}p{0.18\textwidth}p{0.17\textwidth}l}
\hline
Object        & Type   & State & Res.   & Line(s) & Features & Ref.\\
              &        &          & \kms &         &          & \\[1mm]
\hline \\
V616~Mon       & BH     & Q    & 80      &H$\alpha$, $\beta$    &1; 2a; 3&51\\
GU~Mus         & BH     & Q    &         &H$\alpha$             &3       &9\\
V2107~Oph      & BH     & Q    & 120     &H$\alpha$            &1; 3    &25\\
V518~Per       & BH     & O    & \raggedright 100 (H$\beta$); 35 (H$\alpha$)&H$\alpha$, $\beta$, HeII & 1; 2?; 3 &5\\
$''$           & $''$   & Q    & 120     &H$\alpha$            & 1  &26\\
QZ~Vul         & BH     & Q    & 200     &H$\alpha$            &1; 2a   &6\\
$''$           & $''$   & Q    & 120     &H$\alpha$            &1; 2a   &24\\
AR~And         & DN     & Q, O & \raggedright 180 (Q); 130 (O)& \raggedright H$\alpha$ (Q); H$\beta$,$\gamma$ (O) & 1; 7&69 \\
AE~Aqr         & DN     & F    &50 &H$\alpha$        & 2b; 7        &90\\
VY~Aqr         & DN     & Q    &300 &P$\beta$            &1?; 2b     &46\\
OY~Car         & DN     & O    &80  &H$\beta$          &1; 2 or 4; 3 &23\\
AT~Cnc         & DN     & SS   &140 &H$\alpha$         &2; 7         &58\\
EM~Cyg         & DN     & SS   &35  &H$\alpha$         &1; 3        &59\\
SS~Cyg         & DN     & O    &35  &\raggedright H$\alpha$, $\beta$, $\gamma$, HeI, HeII  &1; 3; 4?; 6 &76\\
EX~Dra         & DN     & ?    &80  &H$\alpha$, HeI, HeII           &2ab; 3     &2 \\
$''$           & DN     & Q, O &100 -- 250  & \raggedright H$\alpha$ -- $\delta$, HeI, CII  & 1; 2b&43 \\
$''$           & DN     & O    &35  &\raggedright H$\alpha$,$\beta$, HeI, HeII    &1; 3; 4        &42 \\
U~Gem          & DN     & Q    & 170  & H$\beta$, HeI, HeII &1; 2a; 3  &49, 44 \\
V2051~Oph      & DN     & Q    & 170  &H$\beta$, HeI       &1; 7        &44 \\
IP~Peg         & DN     & Q, O & 150  &H$\beta$, $\gamma$, HeII  &1; 2ab; 3; 4  &50\\
$''$           & $''$   & Q    & 170  &H$\beta$       &1; 7         &44\\
$''$           & $''$   & Q    & 140  &H$\alpha$              &1; 3         &22\\
$''$           & $''$  & O    &35  &H$\alpha$, HeI  &1; 3; 4; 6 &76\\
$''$           & $''$   & O    &35  &H$\alpha$, HeI  &1; 3; 4  &77\\
$''$         & $''$  & Q    &70  &H$\alpha$, $\beta$, $\gamma$ &1; 2a, c; 3  &93, 3\\
$''$           & $''$   & O    &54  &HeII, HeI, MgII              &1; 3; 4; 6 &27, 28\\
$''$           & $''$   & O    &100 &\raggedright HeII, HeI, MgII, CII         &1; 3; 4    &57\\
KT~Per         & DN     &      &200 &H$\beta$                     & 1        &61\\
WZ~Sge         & DN     & Q    & 170& H$\beta$, HeI   &1; 2a     &44\\
$''$           & DN     & Q    & 90  &H$\alpha$       &1; 2a     &75\\
$''$           & DN     & Q    & 300 &P$\beta$        &1; 2b     &46\\
CU~Vel         & DN     & Q    & 320 &H$\alpha$       &1         &55\\
\hline
\end{tabular}\\

See end of Tab.~\ref{tab:part3} for notes.
\label{tab:part1}
\end{table}

\begin{table}
\caption{\index{Doppler map}Doppler maps of CVs and X-ray novae published in
refereed journals as of September 2000.}
\begin{tabular}{p{0.18\textwidth}p{0.1\textwidth}p{0.11\textwidth}p{0.13\textwidth}p{0.18\textwidth}p{0.17\textwidth}l}
\hline
Object        & Type   & State & Res.   & Line(s) & Features & Ref.\\
              &        &          & \kms &         &          & \\[1mm]
\hline \\
GD~552         & DN?    & Q    & 70  &H$\beta$        &1; 2      &35 \\
LY~Hya         & DN     & Q    & 120 &H$\beta$, $\gamma$, $\delta$ &1; 2         &78\\
GP~Com         & IBWD   & --   & 70  &HeI, HeII       &1; 2; 6   &53\\
FO~Aqr         & IP     & --   & 80  &HeII            &2a; 3?    &52\\
$''$           & IP     & --   & 80  &HeII            &         &34$^\dag$ \\
BG~CMi         & IP     & --   &115  &H$\beta$, HeII  &2        &33, 34$^\dag$ \\
PQ~Gem         & IP     & --   &115  &H$\beta$, HeII  &2         &33, 34$^\dag$ \\
EX~Hya         & IP     & --   &170  &H$\beta$         &1; 2a     &44\\
$''$           & IP     & --   &65   &H$\beta$         &         &34$^\dag$ \\
AO~Psc         & IP     & --   &80   &HeII              &         &34$^\dag$ \\
V405~Aur       & IP     & --   &50 &\raggedright H$\alpha$,$\beta$, HeI, HeII  &1; 2b, d &83\\
$''$           & $''$   & --   &120 &H$\alpha$,$\gamma$, HeII  &2b, d   &85\\
RX0757+63      & IP     & --   &280 &H$\beta$         & 1; 2a        &87\\
RX1238-38      & IP     & --   &?   & H$\beta$          &         &34$^\dag$ \\
RX1712-24      & IP     & --   &?   & HeII              &         &34$^\dag$ \\
DQ~Her         & N      & --   &170 &H$\beta$, HeI, HeII &1; 6        &44\\
$''$           & $''$   & --   &120 &\raggedright H$\gamma$, HeI, HeII, CaII & 1; 3 &54\\
BT~Mon         & N      & --   &170 &H$\beta$,$\gamma$, HeII      &2b; 6; 7     &44, 92 \\
$''$           & $''$   & --   &\raggedright 100, blue; 36, red & \raggedright H$\alpha$,$\beta$, HeI, HeII &2b; 6 &74 \\
GQ~Mus         & N      & --   &220 &HeII                    & 1; 3        &16\\
CP~Pup         & N      & --   &170 &H$\beta$, HeII        & 1        &44, 91 \\
PX~And         & NL     & --   &90  & H$\alpha$            & 1; 2b    &31\\
$''$           & $''$   & --   &45  & Balmer, HeII         & 2b, c; 7 &79\\
V1315~Aql      & NL     & --   &65  &H$\beta$, HeII        & 2b; 6; 7    &11\\
$''$           & $''$   & --   &170 &H$\beta$, HeI, HeII   & 2b; 6; 7    &44\\
$''$           & $''$   & --   &115 &H$\beta$, HeII        & 2b; 6; 7    &32\\
UU~Aqr         & NL     & --   &170 &H$\beta$              &1; 2ab    & 44, 45\\
$''$           & $''$   & --   &115 &\raggedright H$\alpha$,$\beta$,$\gamma$, HeI, HeII & 1; 2ab    &38\\
V363~Aur       & NL     & --   &170 &H$\beta$, HeII         &2bc; 7    &44\\
WX~Cen         & NL     & --   &130 &H$\beta$, HeII         &1; 3; 7   &18\\
AC~Cnc         & NL     & --   &170 &H$\beta$, HeII      &2bd; 6; 7          &44\\
V795~Her       & NL     & --   &60  &H$\beta$             &1         &20\\
$''$           & $''$   & --   &70  &\raggedright H$\alpha$,$\beta$,$\gamma$, HeI, HeII &1  &7\\
BH~Lyn         & NL     & L    &75  &H$\beta$,$\gamma$,$\delta$, HeI  & 2b,c; 7  &12\\
$''$           & $''$   & --   &120 &H$\beta$                  &1; 2c  &37\\
\hline
\end{tabular}\\
For codes see Table~\ref{tab:part3}. 

$^\dag$These maps were computed on the spin phase of the \index{white dwarf}white dwarf rather than the
standard orbital phase and thus I have not attempted to describe their features.

\label{tab:part2}
\end{table}

\begin{table}
\caption{\index{Doppler map}Doppler maps of CVs and X-ray novae published in
refereed journals as of September 2000.}
\begin{tabular}{p{0.18\textwidth}p{0.1\textwidth}p{0.11\textwidth}p{0.13\textwidth}p{0.18\textwidth}p{0.17\textwidth}l}
\hline
Object        & Type   & State & Res.   & Line(s) & Features & Ref.\\
              &        &          & \kms &         &          & \\[1mm]
\hline \\
BP~Lyn         & NL     &      &120 &H$\alpha$,$\beta$,$\gamma$  & 1; 2ad  &36\\
$''$           & $''$   &      &80  &H$\alpha$, HeI              & 1; 2bc; 3 &81\\
V347~Pup       & NL     & --    & 120 &Balmer              &1; 3; 4? &82\\
$''$           & $''$   & --    & 120 &H$\beta$            &2ab; 6   &19\\
LX~Ser         & NL     & --     &170 &H$\beta$, HeI, HeII &2a, b; 6    &44\\
SW~Sex         & NL     & --     &170 &H$\beta$, HeI       &1; 2a; 6     &44\\
$''$           & $''$   & --     &75  &\raggedright H$\beta$,$\gamma$,$\delta$, HeI, HeII &1; 2a; 6 &14\\
VZ~Scl         & NL     & --     &170 &H$\beta$            &2ab; 7     &44\\
RW~Tri         & NL     & --     &170 &H$\beta$, HeI, HeII &1, 3     &44\\
$''$           & $''$   & H      &50  &H$\beta$,$\gamma$   &3        &80\\
DW~UMa         & NL     & --     &170 &H$\beta$, HeI, HeII &2b; 6     &44\\
$''$           & $''$   & L      &75  &Balmer              &3         &13\\
UX~UMa         & NL     & --     &170 &H$\beta$, HeII      &1; 6      &44\\
HU~Aqr         & P      & H      &110 &H$\gamma$, HeII     &3; 5      &66, 30\\
AM~Her         & P      & H      &130 &\raggedright NV, SiIV, CIII (\index{ultraviolet}UV) &2b; 3; 7  &21\\
V884~Her       & P      & H      &200 &\raggedright H$\alpha$,$\beta$, HeI, HeII & 2ab  &29\\
BL~Hyi         & P      & H      &160 &H$\alpha$, HeI  & 2a       &56\\
ST~LMi         & P      & --     &70 & NaI, CaII                    & 3        &70\\
$''$           & $''$   & L      &350 & NaI                   &   3      &41\\
V2301~Oph      & P      & H      &80 & \raggedright H$\alpha$,$\beta$,$\gamma$, HeI, HeII  &3; 5?         &71\\
VV~Pup         & P      & H      &90 &H$\alpha$        &3; 5         &17\\
MR~Ser         & P      & --     &70 &NaI, CaII         & 3         &70\\
QQ~Vul         & P      & H      &70 &\raggedright NaI, MgII, HeII, CI    & 3        &10\\
$''$           & $''$   & H      &100 & HeII                    & 3, 5     &68\\
AR~UMa         & P      & M      &100 & \raggedright H$\alpha$,$\beta$, HeI, HeII, MgI, NaI & 3, 5 &65\\
RX0719+65      & P      & H      &300 & H$\beta$, HeII        & 2ab; 7        &86\\
RX1015+09      & P      & H      & ?  & HeII                  &3; 5 &4\\
RX2157+08      & P      & H    &180 & H$\beta$, HeII  & 3      &88\\
\hline
\end{tabular}\\
Notes: 

References are amalgamated when they refer to the same data.

Type codes:  BH = black-hole system; DN = \index{cataclysmic variable!dwarf nova}dwarf nova; 
N = old nova; NL = \index{cataclysmic variable!nova-like}nova-like variable; IBWD = interacting binary \index{white dwarf}white dwarf;
IP = \index{cataclysmic variable!intermediate polar}intermediate polar; P = polar.

State codes (where relevant): Q = \index{quiescence}quiescence; O = \index{outburst}outburst; SS = stand-still; F = flaring; H = high;
L = low; M = middle.

Feature codes: (1) Ring (which may be from a disc), (2) Spot, (3) Secondary star, (4) Spiral shocks,
(5) Gas stream, (6) Low velocity emission, (7) Little structure or low signal-to-noise.

Type (2) = ``spot'' does not necessarily imply stream/disc impact,but just
refers to the appearance of the image. Entries such as 2a refer
to the quadrant the spot is located in (if the orbital phase is known). 
The quadrants start at the upper-left with ``a'', and then go anti-clockwise
from there. A combination such as 2ab implies a spot located on the boundary
of the upper-left and lower-left quadrants.

Type (6) = ``low-velocity emission'' refers to such features as the slingshot
prominences seen in IP~Peg [76] and the emission at low
velocities commonly seen in \index{cataclysmic variable!nova-like}nova-like variables.

\label{tab:part3}
\end{table}

%

\end{document}